# Efficient controlled quantum secure direct communication based on GHZ-like states

Shima Hassanpour [a,*] and Monireh Houshmand [b]

[a] MS Student, Department of Electrical Engineering, Imam Reza International University, Mashhad, Iran
[b] Assistant Professor, Department of Electrical Engineering, Imam Reza International University, Mashhad, Iran

**Abstract**

In this paper, a three-party controlled quantum secure direct communication protocol based on GHZ-like state is proposed. In this scheme, the receiver can obtain the sender's two-secret bits under the permission of the controller. By using entanglement swapping, no qubits carrying secret messages are transmitted. Therefore, if the perfect quantum channel is used, the protocol is completely secure. The motivation behind utilizing GHZ-like state as a quantum channel is that if a qubit is lost in the GHZ-like state the other two qubits are still entangled. The proposed protocol improves the efficiency of the previous ones.

*Key words: quantum cryptography; controlled quantum secure direct communication; entanglement swapping; GHZ-like state; efficiency*

## 1. Introduction

Quantum cryptography [1], is a branch of quantum information science which prepares an absolute way to ensure the security of communication. Quantum cryptography began by discovering of quantum key distribution (QKD). The first QKD protocol (BB84) was proposed in 1984 [2], by Bennett and Brassard. Quantum cryptography developed so rapidly that nowadays other branches of this field exist. They can be categorized into quantum secure direct communication (QSDC) [3 − 24], quantum secret sharing (QSS) [25 − 30], quantum state sharing (QSTS) [31 − 33], deterministic secure quantum communication (DSQC) [34 − 36] and so on.

In contrast to QKD [37,38], QSDC permits the secret message to be transmitted directly without first sharing a private key. The works on this research field attracted a great deal of attention and can be divided into two groups, one utilizes single photons [3 − 5] the other is based on entangled states [6]. Many kinds of channels in QSDC have been applied such as Bell state [6,7], GHZ state [12 − 14], $W$ state [15 − 19], GHZ-like state [20 − 24] and so on.

The first QSDC protocol was proposed by Beige et al., in 2002 [3]. The scheme is based on single photons. In the same year, Boström and Felbinger [8] presented a ping-pong QSDC scheme by using EPR pairs. However, their scheme is not secure in a noisy quantum channel [9]. One year later, Deng et al., [10] presented a two-step QSDC scheme, using EPR pairs. In all these QSDC schemes, the qubits carrying the secret message are sent in a quantum channel. So, Eve can have a successful attack and she is able to steal some part of the secret message. To solve this problem, Yan et al., [11] presented a QSDC scheme based on EPR pairs and teleportation. In this scheme, the receiver can find out the secret message with the help of sender's classical information. Also, no qubits carrying secret message are transmitted. Therefore, this protocol is unconditionally secure, if the perfect quantum channel is used.

Recently, a new type of QSDC, controlled quantum secure direct communication (CQSDC), has attracted lots of attention. In CQSDC scheme, a sender can communicate with the receiver under the permission of a controller. In fact, without the controller's admission, the receiver cannot get any part of the secret message. In 2005 [12],

---

* Corresponding author at: Department of Electrical Engineering, Imam Reza International University, Mashhad, Iran Tel.: +98-882-179-9.

E-mail address: shimahassanpour@yahoo.com.

Xia et al., presented a CQSDC scheme by using GHZ entangled state via swapping quantum entanglement. Another quantum channel that has been applied nowadays is *W* states [15].

In recent years, some researchers have worked on QSDC in which GHZ-like states are used as a quantum channel. The most important property of GHZ-like state is that its entanglement is maximally robust under disposal of any one of the three qubits [39]. This new branch of QSDC needs lots of attentions in order to be developed. GHZ-like state can be produced by a single qubit state, an EPR state and a CNOT operation. We can demonstrate them as Eq. (1):

$$
\begin{aligned}
&|\xi_{000}\rangle_{123} = \tfrac{1}{\sqrt{2}}(|\Psi^+ 0\rangle + |\Psi^- 1\rangle)_{123}, &\quad &|\xi_{001}\rangle_{123} = \tfrac{1}{\sqrt{2}}(|\Psi^+ 0\rangle - |\Psi^- 1\rangle)_{123}, \\
&|\xi_{010}\rangle_{123} = \tfrac{1}{\sqrt{2}}(|\Psi^- 0\rangle - |\Psi^+ 1\rangle)_{123}, &\quad &|\xi_{011}\rangle_{123} = \tfrac{1}{\sqrt{2}}(|\phi^+ 1\rangle - |\phi^- 0\rangle)_{123}, \\
&|\xi_{100}\rangle_{123} = \tfrac{1}{\sqrt{2}}(|\phi^+ 0\rangle + |\phi^- 1\rangle)_{123}, &\quad &|\xi_{101}\rangle_{123} = \tfrac{1}{\sqrt{2}}(|\phi^+ 0\rangle - |\phi^- 1\rangle)_{123}, \\
&|\xi_{110}\rangle_{123} = \tfrac{1}{\sqrt{2}}(|\phi^- 0\rangle + |\phi^+ 1\rangle)_{123}, &\quad &|\xi_{111}\rangle_{123} = \tfrac{1}{\sqrt{2}}(|\Psi^+ 1\rangle - |\Psi^- 0\rangle)_{123}.
\end{aligned} \quad (1)
$$

In 2005, Gao et al., [20] proposed a CQSDC scheme based on three-qubit GHZ-like state as a quantum channel. In 2011, Dong et al., [21] proposed another CQSDC scheme based on GHZ-like state and imperfect Bell state measurement. Also, they utilize the controlled quantum teleportation technique. In their scheme, the receiver can get one-bit secret message from transmitter under the permission of a controller. One year later, Banerjee and Pathak [22] proposed a QSDC scheme in which a sender transmits three secret messages to a receiver. Also, in this protocol the qubits carrying the secret messages are transmitted in a quantum channel. In that year, Kao and Hwang [23] proved that in Gao et al.'s scheme [20], the receiver can reveal the whole secret message without the controller's permission. Also, in 2013, an improved scheme [24] was presented by Kao and Hwang to avoid this flaw.

Based on the GHZ-like state property, many researchers utilized it as a quantum channel [29,40 − 42], in this paper, a CQSDC protocol is proposed. In this scheme, three-particle GHZ-like states are produced by a sender, Alice. The second particle of each GHZ-like state is transmitted to her agent, Bob and the third particle is transmitted to a controller, Charlie. After preparing the quantum channel, Alice encodes her two-bit secret messages, by applying pre-defined unitary operations. Then, Bob can get the secret messages with his measurement's result and classical bits that received from Alice and Charlie. Also, we did not use teleportation technique. But, we utilize entanglement swapping for transmitting two-bit secret messages. As a result, the proposed scheme is unconditionally secure if perfect quantum channel is applied. The results will show our scheme improves the efficiency of the previous work.

The rest of this paper is organized as follows. A brief review on CQSDC schemes which utilize GHZ-like state as a quantum channel is presented in Section 2. In Section 3, proposed CQSDC protocol is presented in details. The security analyses are investigated in Section 4. In Section 5, the comparison and efficiency analyses are given. Finally, the conclusions are provided in Section 6.

## 2. Previous CQSDC schemes based on GHZ-like state

In this section, previous CQSDC protocols which utilize GHZ-like state as a quantum channel are reviewed. First, it is essential to introduce some important basis measurements that are being applied in QSDC protocols. The *Z*-basis is $\{|0\rangle, |1\rangle\}$. Also, the *X*-basis is $\{|+\rangle, |-\rangle\}$, where $|+\rangle, |-\rangle$ are defined as Eq. (2),

$$|+\rangle = \frac{1}{\sqrt{2}}(|0\rangle + |1\rangle), \qquad |-\rangle = \frac{1}{\sqrt{2}}(|0\rangle - |1\rangle). \tag{2}$$

The Bell states are denoted as Eq. (3),

$$|\phi^{\pm}\rangle_{12} = \frac{1}{\sqrt{2}}(|00\rangle \pm |11\rangle)_{12}, \qquad |\psi^{\pm}\rangle_{12} = \frac{1}{\sqrt{2}}(|01\rangle \pm |10\rangle)_{12}. \tag{3}$$

### 2.1. Dong et al. scheme

This protocol consists of four steps as follows:

Step 1 The controller, provides *N* three-particle GHZ-like states as Eq. (4),

$$|\xi\rangle_{123} = \tfrac{1}{2}(|000\rangle + |110\rangle + |011\rangle + |101\rangle)_{123}. \tag{4}$$

He sends particles one and two of each GHZ-like state as *A* and *B* sequences to the sender and receiver and keeps the third particles as *C* sequence for himself.

Step 2 For checking the security of quantum channel, the receiver randomly selects *Z*-basis or *X*-basis measurements. Then, he performs it on his particles that he picked them out for eavesdropping check. Also, the sender and the controller apply the same measurement basis on the corresponding particles that the receiver announced their positions. So, by comparing their results, the communicators can evaluate the error rate.

Step 3 If the quantum channel is secure, the controller measures his particle in the *Z*-basis and announces his results publicly. So, the two communicators can start their communication. For sending secret message, the sender produces extra qubit; $|0\rangle$ for "0" and $|1\rangle$ for "1". Then, she measures her two particles on Bell basis measurement and announces her result as a classical bit to the receiver.

Step 4 After the receiver gets the classical bits, he applies *Z*-basis measurement on his particle. By comparing his results and the sender's classical bits, he can find out the secret message.

*2.2. Kao et al. scheme*

This protocol is implemented by the following steps:

Step 1 The controller provides a sequence of three particle GHZ-like state as Eq. (4). He sends particles two and three of each GHZ-like state for Alice and Bob as a transmitter and receiver. Also, he maintains the first particles for himself.

Step 2 For eavesdropping check, three legitimate users pick up randomly enough particles for eavesdropping process. Then, they apply appropriate measurement basis and discuss publicly about the entanglement relation between these particles. So, by checking their outcomes they can evaluate the error rate.

Step 3 If the channel is secure, the sender encodes her secret message. She creates an extra particle depending on her secret message that she wants to transmit. If her secret message is "0" ("1"), she produces $|0\rangle(|1\rangle)$. Then, she performs Bell measurement on her two particles and transmits her result as classical bits to the receiver.

Step 4 In this step, the receiver applies unitary operations according to the classical bits received from the sender and the controller. Finally, by performing *X*- basis measurement he can find out the secret message.

## 3. The proposed QSDC protocol

In the proposed protocol, we suppose that Alice attempts to transmit her secret messages to Bob under the control of Charlie. The scheme can be achieved in four steps. The first step is the preparation in which a quantum channel is prepared. The second step is security checking that the security of the quantum channel is discussed. The third step is secret communication where the sender encodes her secret messages. Finally, the decoding process is discussed in the last step which is called message extraction. Let us start by illustrating entanglement swapping used in the first step. Then the details of the proposed protocol are discussed.

*Entanglement swapping*

Entanglement swapping [43] is one of the most fascinating aspect of quantum communication. This method can entangle two particles which have never interacted. This feature is illustrated with two EPR pairs in which particles one and two are entangled together and particles three and four are entangled together. Also, the sender has particles one and four and the receiver has particles two and three in their possessions. In this case, if the sender performs a Bell state measurement on her particles, the state collapses and the receiver's particles become entangled with each other as follows:

$$|\emptyset^+\rangle_{12} \otimes |\emptyset^+\rangle_{34} = \frac{1}{\sqrt{2}}(|00\rangle + |11\rangle)_{12} \otimes \frac{1}{\sqrt{2}}(|00\rangle + |11\rangle)_{34}$$
$$= \frac{1}{2}(|00\rangle_{12}|00\rangle_{34} + |00\rangle_{12}|11\rangle_{34} + |11\rangle_{12}|00\rangle_{34} + |11\rangle_{12}|11\rangle_{34})$$
$$= \frac{1}{2}(|00\rangle_{14}|00\rangle_{23} + |01\rangle_{14}|01\rangle_{23} + |10\rangle_{14}|10\rangle_{23} + |11\rangle_{14}|11\rangle_{23})$$
$$= \frac{1}{2}(|\emptyset^+\rangle_{14}|\emptyset^+\rangle_{23} + |\emptyset^-\rangle_{14}|\emptyset^-\rangle_{23} + |\psi^+\rangle_{14}|\psi^+\rangle_{23} + |\psi^-\rangle_{14}|\psi^-\rangle_{23}). \quad (5)$$

According to Eq. (5), if the sender's result is $|\emptyset^+\rangle_{14}$, the receiver's outcome will be $|\emptyset^+\rangle_{23}$ that are maximally entangled states. Also, similar entanglement swapping can be utilized in other Bell states.

## 3.1. Preparation phase

In this scheme Alice produces large enough number ($2N$) of three-particle GHZ-like state which is denoted as Eq. (6),

$$|P\rangle_{a_i b_i c_i} = \frac{1}{2}(|100\rangle + |010\rangle + |001\rangle + |111\rangle)_{a_i b_i c_i}, \tag{6}$$

where a, b, c represent the three qubits in a tripartite GHZ-like state and $i \in \{1, 2N\}$. These GHZ-like states can be defined in $2N$ groups as: $[P(a_1, b_1, c_1), P(a_2, b_2, c_2), \ldots, P(a_i, b_i, c_i), \ldots P(a_{2N}, b_{2N}, c_{2N})]$. Alice divides the series of GHZ-like states into three sequences. $A, B$ and $C$ sequences are composed of the first, the second and the third particles respectively.

$A: [P(a_1), P(a_2), \ldots P(a_{2N})]$
$B: [P(b_1), P(b_2), \ldots P(b_{2N})]$
$C: [P(c_1), P(c_2), \ldots P(c_{2N})]$

Alice divides $A$ sequence into many groups. Each group is composed of two particles, particles $P(a_1)$ and $P(a_2)$ as group one, particles $P(a_3)$ and $P(a_4)$ as group two, etc. Then, she sends $B$ and $C$ sequences to Bob and Charlie respectively. Also, Bob and Charlie do the same as Alice for storing their particles.

After Bob and Charlie affirm that they have received the sequences, Alice chooses randomly some groups for encoding her secret messages. The remaining groups are taken as checking groups for ensuring the security of the quantum channel.

## 3.2. Security checking phase

Step 1 Alice randomly chooses one of two measuring bases, $X$-basis or $Z$-basis, to measure her particles. After that, she announces the order of the particles, the measuring basis and the results of her measurement. The GHZ-like state can be written in $X$-basis as Eq. (7),

$$|P\rangle_{a_i b_i c_i} = \frac{1}{\sqrt{2}}(|+++\rangle - |---\rangle)_{a_i b_i c_i}. \tag{7}$$

Step 2 Bob and Charlie perform their measurements under the same measuring basis on each of the corresponding particles.

Step 3 They compare their measurement's results. According to Table 1, measurement results of the users should be correlated, if there is no eavesdropper. If the error rate is more than the threshold, they abort the protocol; otherwise, they execute the next state.

**Table 1**. The measurement bases and corresponding measurement results.

| Basis | Alice's state | Bob's state | Charlie's state |
|---|---|---|---|
| $\sigma_x$ | $|+\rangle$ | $|+\rangle$ | $|+\rangle$ |
| | $|-\rangle$ | $|-\rangle$ | $|-\rangle$ |
| $\sigma_z$ | $|1\rangle$ | $|0\rangle$ | $|0\rangle$ |
| | $|0\rangle$ | $|1\rangle$ | $|0\rangle$ |
| | $|0\rangle$ | $|0\rangle$ | $|1\rangle$ |
| | $|1\rangle$ | $|1\rangle$ | $|1\rangle$ |

## 3.3. Secret communication phase

Step 1 After ensuring the security of quantum channel, Alice encodes her secret messages on the remaining groups of $A$ sequence. She applies the Pauli operations defined in Table 2 on her first particle in each group. The Pauli operations are defined as Eq. (8):

**Table 2.** Defined operations for encoding two bits secret messages.

| Secret Message | Operations |
|---|---|
| 00 | $U_0 \otimes I$ |
| 01 | $U_1 \otimes I$ |
| 10 | $U_2 \otimes I$ |
| 11 | $U_3 \otimes I$ |

$$U_0 = I = |0\rangle\langle 0| + |1\rangle\langle 1|,$$
$$U_1 = \sigma_x = |0\rangle\langle 1| + |1\rangle\langle 0|,$$
$$U_2 = i\sigma_y = |0\rangle\langle 1| - |1\rangle\langle 0|,$$
$$U_3 = \sigma_z = |0\rangle\langle 0| - |1\rangle\langle 1|. \quad (8)$$

As an example, if Alice wants to send "00", she performs $U_0$ operation on the first particle of group one. The state of the whole system which is prepared of particles $(a_1, b_1; a_2, b_2)$, will be as Eq. (9):

$$\begin{aligned}|P_{00}\rangle &= (U_0 \otimes I^{\otimes 5})(|P\rangle_{a_1 b_1 c_1} \otimes |P\rangle_{a_2 b_2 c_2}) \\ &= \tfrac{1}{4}[(|1100\rangle + |1001\rangle + |0110\rangle + |0011\rangle)_{a_1 a_2 b_1 b_2}|00\rangle_{c_1 c_2} \\ &\quad + (|0000\rangle + |0101\rangle + |1010\rangle + |1111\rangle)_{a_1 a_2 b_1 b_2}|11\rangle_{c_1 c_2} \\ &\quad + (|1000\rangle + |1101\rangle + |0010\rangle + |0111\rangle)_{a_1 a_2 b_1 b_2}|01\rangle_{c_1 c_2} \\ &\quad + (|0100\rangle + |0001\rangle + |1110\rangle + |1011\rangle)_{a_1 a_2 b_1 b_2}|10\rangle_{c_1 c_2}] \\ &= \tfrac{1}{2\sqrt{2}}[(|\emptyset^+\rangle_{a_1 a_2}|\emptyset^+\rangle_{b_1 b_2} + |\Psi^+\rangle_{a_1 a_2}|\Psi^+\rangle_{b_1 b_2})|\emptyset^+\rangle_{c_1 c_2} \\ &\quad -(|\emptyset^-\rangle_{a_1 a_2}|\emptyset^-\rangle_{b_1 b_2} + |\Psi^-\rangle_{a_1 a_2}|\Psi^-\rangle_{b_1 b_2})|\emptyset^-\rangle_{c_1 c_2} \\ &\quad -(|\Psi^+\rangle_{a_1 a_2}|\emptyset^+\rangle_{b_1 b_2} + |\emptyset^+\rangle_{a_1 a_2}|\Psi^+\rangle_{b_1 b_2})|\Psi^+\rangle_{c_1 c_2} \\ &\quad -(|\Psi^-\rangle_{a_1 a_2}|\emptyset^-\rangle_{b_1 b_2} + |\emptyset^-\rangle_{a_1 a_2}|\Psi^-\rangle_{b_1 b_2})|\Psi^-\rangle_{c_1 c_2}]. \end{aligned} \quad (9)$$

As we can see in Eq. (9), the entanglement swapping has been applied.

Step 2 Alice asks Charlie to perform Hadamard operation on his two particles in a group. The Hadamard matrix $(H)$ is defined in Eq. (10),

$$H = \tfrac{1}{\sqrt{2}}(|0\rangle\langle 0| - |1\rangle\langle 1| + |0\rangle\langle 1| + |1\rangle\langle 0|). \quad (10)$$

According to Eq. (10), the state is changed into $|P'_{00}\rangle$ as follows:

$$\begin{aligned}|P'_{00}\rangle &= \tfrac{1}{4}[(|\emptyset^+\rangle_{a_1 a_2}|\emptyset^+\rangle_{b_1 b_2} + |\Psi^+\rangle_{a_1 a_2}|\Psi^+\rangle_{b_1 b_2} + |\Psi^-\rangle_{a_1 a_2}|\emptyset^-\rangle_{b_1 b_2} + |\emptyset^-\rangle_{a_1 a_2}|\Psi^-\rangle_{b_1 b_2})|\emptyset^-\rangle_{c_1 c_2} \\ &\quad + (|\emptyset^+\rangle_{a_1 a_2}|\emptyset^+\rangle_{b_1 b_2} + |\Psi^+\rangle_{a_1 a_2}|\Psi^+\rangle_{b_1 b_2} - |\Psi^-\rangle_{a_1 a_2}|\emptyset^-\rangle_{b_1 b_2} - |\emptyset^-\rangle_{a_1 a_2}|\Psi^-\rangle_{b_1 b_2})|\Psi^+\rangle_{c_1 c_2} \\ &\quad + (|\emptyset^-\rangle_{a_1 a_2}|\emptyset^-\rangle_{b_1 b_2} + |\Psi^-\rangle_{a_1 a_2}|\Psi^-\rangle_{b_1 b_2} + |\Psi^+\rangle_{a_1 a_2}|\emptyset^+\rangle_{b_1 b_2} + |\emptyset^+\rangle_{a_1 a_2}|\Psi^+\rangle_{b_1 b_2})|\Psi^-\rangle_{c_1 c_2} \\ &\quad + (-|\emptyset^-\rangle_{a_1 a_2}|\emptyset^-\rangle_{b_1 b_2} - |\Psi^-\rangle_{a_1 a_2}|\Psi^-\rangle_{b_1 b_2} + |\Psi^+\rangle_{a_1 a_2}|\emptyset^+\rangle_{b_1 b_2} + |\emptyset^+\rangle_{a_1 a_2}|\Psi^+\rangle_{b_1 b_2})|\emptyset^+\rangle_{c_1 c_2}]. \end{aligned} \quad (11)$$

After Charlie performs Hadamard operation, he measures his particles in a group in Bell states. If the measurement result is $|\emptyset^-\rangle$ or $|\Psi^+\rangle$, he sends "0" as a classical bit to Bob; if the measurement outcome is $|\emptyset^+\rangle$ or $|\Psi^-\rangle$, he sends "1" to Bob.

Step 3 After Alice finds out that Charlie finished his work, she measures her particles in Bell states and transmits a classical bit to Bob as Charlie did.

### 3.4. Message extraction phase

Step 1 Bob apply exclusive-or gate of a classical bit with Charlie's information. If Bob receives "00" or "01", he considers "0" and if he receives "10" or "11", he considers "1" as the classical bit. Then, he performs suitable unitary operation as Table 3 shows.

Step 2 After performing unitary operation, he measures his particles in Bell states. Now, Bob can infer the secret messages that Alice has sent to him. Actually, he compares his measurement's result with Alice's outcome. The results of this communication are shown in Table 3.

**Table 3**. Relation between the measurement results, classical information, Bob's operation and the secret message.

| Alice's Result | Charlie's Result | Classical Information A to B | Classical Information C to B | Bob's Unitary operation | Bob's Result | Secret Message |
|---|---|---|---|---|---|---|
| $|\phi^+\rangle$ | $|\phi^-\rangle, |\psi^+\rangle$ | 00 | 0 | $I$ | $|\phi^+\rangle$ | |
| $|\phi^+\rangle$ | $|\phi^+\rangle, |\psi^-\rangle$ | 00 | 1 | $\delta_x$ | $|\phi^+\rangle$ | |
| $|\psi^+\rangle$ | $|\phi^-\rangle, |\psi^+\rangle$ | 01 | 0 | $I$ | $|\psi^+\rangle$ | |
| $|\psi^+\rangle$ | $|\phi^+\rangle, |\psi^-\rangle$ | 01 | 1 | $\delta_x$ | $|\psi^+\rangle$ | 00 |
| $|\phi^-\rangle$ | $|\phi^-\rangle, |\psi^+\rangle$ | 10 | 0 | $\delta_x$ | $|\phi^-\rangle$ | |
| $|\phi^-\rangle$ | $|\phi^+\rangle, |\psi^-\rangle$ | 10 | 1 | $I$ | $|\phi^-\rangle$ | |
| $|\psi^-\rangle$ | $|\phi^-\rangle, |\psi^+\rangle$ | 11 | 0 | $\delta_x$ | $|\psi^-\rangle$ | |
| $|\psi^-\rangle$ | $|\phi^+\rangle, |\psi^-\rangle$ | 11 | 1 | $I$ | $|\psi^-\rangle$ | |
| $|\phi^+\rangle$ | $|\phi^-\rangle, |\psi^+\rangle$ | 00 | 0 | $I$ | $|\psi^+\rangle$ | |
| $|\phi^+\rangle$ | $|\phi^+\rangle, |\psi^-\rangle$ | 00 | 1 | $\delta_x$ | $|\psi^+\rangle$ | |
| $|\psi^+\rangle$ | $|\phi^-\rangle, |\psi^+\rangle$ | 01 | 0 | $I$ | $|\phi^+\rangle$ | |
| $|\psi^+\rangle$ | $|\phi^+\rangle, |\psi^-\rangle$ | 01 | 1 | $\delta_x$ | $|\phi^+\rangle$ | 01 |
| $|\phi^-\rangle$ | $|\phi^-\rangle, |\psi^+\rangle$ | 10 | 0 | $\delta_x$ | $|\psi^-\rangle$ | |
| $|\phi^-\rangle$ | $|\phi^+\rangle, |\psi^-\rangle$ | 10 | 1 | $I$ | $|\psi^-\rangle$ | |
| $|\psi^-\rangle$ | $|\phi^-\rangle, |\psi^+\rangle$ | 11 | 0 | $\delta_x$ | $|\phi^-\rangle$ | |
| $|\psi^-\rangle$ | $|\phi^+\rangle, |\psi^-\rangle$ | 11 | 1 | $I$ | $|\phi^-\rangle$ | |
| $|\phi^+\rangle$ | $|\phi^-\rangle, |\psi^+\rangle$ | 00 | 0 | $I$ | $|\phi^-\rangle$ | |
| $|\phi^+\rangle$ | $|\phi^+\rangle, |\psi^-\rangle$ | 00 | 1 | $\delta_x$ | $|\phi^-\rangle$ | |
| $|\psi^+\rangle$ | $|\phi^-\rangle, |\psi^+\rangle$ | 01 | 0 | $I$ | $|\psi^-\rangle$ | |
| $|\psi^+\rangle$ | $|\phi^+\rangle, |\psi^-\rangle$ | 01 | 1 | $\delta_x$ | $|\psi^-\rangle$ | 10 |
| $|\phi^-\rangle$ | $|\phi^-\rangle, |\psi^+\rangle$ | 10 | 0 | $\delta_x$ | $|\phi^+\rangle$ | |
| $|\phi^-\rangle$ | $|\phi^+\rangle, |\psi^-\rangle$ | 10 | 1 | $I$ | $|\phi^+\rangle$ | |
| $|\psi^-\rangle$ | $|\phi^-\rangle, |\psi^+\rangle$ | 11 | 0 | $\delta_x$ | $|\psi^+\rangle$ | |
| $|\psi^-\rangle$ | $|\phi^+\rangle, |\psi^-\rangle$ | 11 | 1 | $I$ | $|\psi^+\rangle$ | |
| $|\phi^+\rangle$ | $|\phi^-\rangle, |\psi^+\rangle$ | 00 | 0 | $I$ | $|\psi^-\rangle$ | |
| $|\phi^+\rangle$ | $|\phi^+\rangle, |\psi^-\rangle$ | 00 | 1 | $\delta_x$ | $|\psi^-\rangle$ | |
| $|\psi^+\rangle$ | $|\phi^-\rangle, |\psi^+\rangle$ | 01 | 0 | $I$ | $|\phi^-\rangle$ | |
| $|\psi^+\rangle$ | $|\phi^+\rangle, |\psi^-\rangle$ | 01 | 1 | $\delta_x$ | $|\phi^-\rangle$ | 11 |
| $|\phi^-\rangle$ | $|\phi^-\rangle, |\psi^+\rangle$ | 10 | 0 | $\delta_x$ | $|\psi^+\rangle$ | |
| $|\phi^-\rangle$ | $|\phi^+\rangle, |\psi^-\rangle$ | 10 | 1 | $I$ | $|\psi^+\rangle$ | |
| $|\psi^-\rangle$ | $|\phi^-\rangle, |\psi^+\rangle$ | 11 | 0 | $\delta_x$ | $|\phi^+\rangle$ | |
| $|\psi^-\rangle$ | $|\phi^+\rangle, |\psi^-\rangle$ | 11 | 1 | $I$ | $|\phi^+\rangle$ | |

## 4. Security

In the proposed scheme, because of using entanglement swapping, no qubits carrying secret messages are transmitted in a quantum channel. Also, only in the preparation phase Eve may have an attack. So, if the quantum channel between Alice, Bob and Charlie is secure, the protocol will be unconditionally secure.

In this section, three important security phases are discussed. In the first phase, the security of the scheme without the controller's permission is investigated. In the second phase, the security of the proposed scheme under the control of Eve is studied. Finally, in the third phase, two kinds of attacks that Eve may use them in the preparation phase are analyzed.

*4.1. Security analysis without the controller's permission*

One of the most important concerns in the design and analysis of CQSDC protocols is that the receiver cannot obtain any part of the secret message without the permission of the controller. In the proposed protocol as we can see in Table 3, the receiver cannot get any part of the secret message without the controller's classical information. As an example, if Alice's classical information sending through classical channel is "00", Bob cannot find out the secret message correctly. According to Eq. (12), all of the secret messages have the same probability.

$$prob(xy = 00|kl) = prob(xy = 01|kl) = prob(xy = 10|kl) = prob(xy = 11|kl) = \frac{1}{4}, \quad (12)$$
$$k, l \in \{(00), (01), (10), (11)\}$$

where $xy, kl$ indicate the secret message and Alice's classical information respectively.

*4.2. Security analysis by displaying all of the classical information*

According to Table 3, if all of the classical information is displayed, Eve cannot get any part of the secret bits. We can demonstrate it as follows:

$$prob(xy = 00|klm) = prob(xy = 01|klm) = prob(xy = 10|klm) = prob(xy = 11|klm) = \frac{1}{4}, \quad (13)$$
$$k, l, m \in \{(000), (001), (010), (011), (100), (101), (110), (111)\}$$

where $xy, kl, m$ indicate the secret message and Alice's and Charlie's classical information respectively. According to Eq. (13), Eve cannot distinguish the secret bits correctly. In fact, because of appearing all the eight possible classical information for each two-bit secret message, she cannot get any part of the secret message.

*4.3. Types of Eve's attacks in a quantum channel*

In this section, three kinds of attacks that Eve may utilize in the preparation phase are analyzed [44,45]. The first one is the intercept- and- resend attack, the second one is the controlled- not attack and the third one is the entanglement-and-measure attack. These attacks are described as follows:

*4.3.1. Intercept-and-resend attack*

In this attack, Eve captures and measures the transmitting particles in $Z$ or $X$ basis randomly and sends the replacement qubit prepared in the state she measured to Alice (or Charlie). In this action, Eve cannot get any useful information. In fact, because of no correlations between Alice's (or Charlie's) particles and the fake ones, Eve's attack will be found during the security checking phase. Suppose Eve's measurement basis is $X$. In the security checking phase, if the users choose $X$-basis, according to Eq. (7) Eve's attack would not be detected and when the users basis is $Z$ as described in Eq. (14), her existence would be detected with $1/2$ probability.

$$\frac{1}{2}(|100\rangle + |010\rangle + |001\rangle + |111\rangle) = \frac{1}{\sqrt{2}}(|1+0\rangle + |0+0\rangle + |0+1\rangle + |1+1\rangle$$
$$+|1-0\rangle - |0-0\rangle + |0-1\rangle - |1-1\rangle). \quad (14)$$

Also, suppose Eve's measurement basis is $Z$. If the users choose $Z$ basis, According to Eq. (6) Eve's interface would not be detected and when the users basis is $X$ as demonstrated in Eq. (15), her existence would be detected with $1/2$ probability.

$$\frac{1}{\sqrt{2}}(|+++\rangle - |---\rangle) = \frac{1}{2}(|+0+\rangle - |-0-\rangle + |+1+\rangle + |-1-\rangle). \quad (15)$$

*4.3.2. Controlled-not attack*

In this attack, Eve intercepts the qubit sent from Bob to Alice (or Charlie) and then she performs controlled-not gate to add an additional qubit on every GHZ-like state. She wants to find out the secret message by measuring these additional qubits. Here we find out the error rate where Alice's (Charlie's) qubit is controller and Eve's qubit is target. The state of the whole system after Eve's operation will be as Eq. (16),

$$CNOT_{aE} = \left[\frac{1}{2}(|100\rangle + |010\rangle + |001\rangle + |111\rangle)_{abc}(a|0\rangle + b|1\rangle)_E\right]$$
$$= \frac{1}{2}(a|1000\rangle + a|0100\rangle + a|0010\rangle + a|1110\rangle + b|0001\rangle + b|1101\rangle + b|1011\rangle + b|0111\rangle)_{abcE}. \quad (16)$$

After measuring the qubits in the security checking phase, the Eve's existence can be detected with probabilities of $\frac{3}{4}\{|a-b|^2\}$ and $|b|^2$, if the measurement basis is $X$ and $Z$ respectively.

### 4.3.3. Entanglement-and-measure attack

In this kind of attack, Eve produces auxiliary particles $|E_i\rangle$, and entangles them with a particle a (or c) before Alice (or Charlie) received it. Also, Eve uses unitary operation $P_{NE}$, on the pair of particles a (or c) and her particle $E$. We can describe it as Eqs. (17 − 20),

$$P_{NE}.|0\rangle|E_i\rangle = p_{00}|0\rangle|e_{00}\rangle + p_{01}|1\rangle|e_{01}\rangle, \quad (17)$$
$$P_{NE}.|1\rangle|E_i\rangle = p_{10}|0\rangle|e_{10}\rangle + p_{11}|1\rangle|e_{11}\rangle, \quad (18)$$
$$P_{NE}.|+\rangle|E_i\rangle = \tfrac{1}{2}[|+\rangle(p_{00}|e_{00}\rangle + p_{01}|e_{01}\rangle + p_{10}|e_{10}\rangle + p_{11}|e_{11}\rangle) + |-\rangle(p_{00}|e_{00}\rangle - p_{01}|e_{01}\rangle + p_{10}|e_{10}\rangle - p_{11}|e_{11}\rangle)], (19)$$
$$P_{NE}.|-\rangle|E_i\rangle = \tfrac{1}{2}[|+\rangle(p_{00}|e_{00}\rangle + p_{01}|e_{01}\rangle - p_{10}|e_{10}\rangle - p_{11}|e_{11}\rangle) + |-\rangle(p_{00}|e_{00}\rangle - p_{01}|e_{01}\rangle - p_{10}|e_{10}\rangle + p_{11}|e_{11}\rangle)], (20)$$

where $|e_{ij}\rangle$, and $N$ indicate the state of $|E_i\rangle$, and Alice's (or Charlie's) particles respectively. The unitary operation is describe as Eq. (21),

$$P_{NE} = \begin{bmatrix} p_{00} & p_{01} \\ p_{10} & p_{11} \end{bmatrix}. \quad (21)$$

After Eve performed unitary operation, she resends the particle $N$, to the legal receivers. The state of the whole system will be as Eq. (22),

$$|\emptyset\rangle = \tfrac{1}{2}[p_{00}(|100\rangle + |001\rangle)|e_{00}\rangle + p_{01}(|110\rangle + |011\rangle)|e_{01}\rangle + p_{10}(|000\rangle + |101\rangle)|e_{10}\rangle + p_{11}(|010\rangle + |111\rangle)|e_{11}\rangle$$
$$= \tfrac{1}{2\sqrt{2}}[|+++\rangle(p_{00}|e_{00}\rangle + p_{01}|e_{01}\rangle + p_{10}|e_{10}\rangle + p_{11}|e_{11}\rangle) + |+-+\rangle(p_{00}|e_{00}\rangle - p_{01}|e_{01}\rangle + p_{10}|e_{10}\rangle - p_{11}|e_{11}\rangle)$$
$$-|+-\rangle(p_{00}|e_{00}\rangle + p_{01}|e_{01}\rangle - p_{10}|e_{10}\rangle - p_{11}|e_{11}\rangle) - |---\rangle(p_{00}|e_{00}\rangle - p_{01}|e_{01}\rangle - p_{10}|e_{10}\rangle + p_{11}|e_{11}\rangle)). \quad (22)$$

According to Eq. (22) and Table 1, the Eve's existence can be detected in the security checking process with probabilities of $\tfrac{1}{2}\{|p_{01}|^2 + |p_{10}|^2\}$ and $\tfrac{1}{4}\{|p_{00}|^2 + |p_{01}|^2 + |p_{10}|^2 + |p_{11}|^2\}$ if the user's measurement basis is $Z$ and $X$, respectively.

## 5. Comparison

In this scheme, we use the idea of entanglement swapping to transmit secret messages. In contrast to the scheme of Banerjee et al., [22], no qubits carrying the secret messages are transmitted in the proposed protocol. One of the disadvantages of transmitting the qubits that carry the secret messages is, the security of the quantum channel must be checked two times. Also, in the both schemes of Dong et al., [21], and Kao et al., [23], teleportation techniques have been utilized. Actually, they produce extra qubits for encoding one- bit secret messages. But, in this paper, by using entanglement swapping two-bit secret messages have been transmitted to the receiver.

In addition, the important factor that can be applied for comparing the performance of protocols is efficiency [34,46]. In quantum communication protocols, it is defined as Eq. (23),

$$\eta_1 = \frac{m_u}{q_k + b_k}, \quad (23)$$

where $m_u$, $q_k$ and $b_k$ indicate the number of secret bits, the number of total qubits and the number of classical bits respectively. Since transmission of qubits is more complex and expensive than the classical bits, we can redefine the efficiency as Eq. (24),

$$\eta_2 = \frac{m_u}{q_k}. \quad (24)$$

Table 4 shows the efficiency comparison of the proposed protocol with the previous schemes. It is obvious that the efficiency of the proposed protocol is higher than Dong et al.'s and Kao et al.'s in both definitions.

Table 4. Efficiency comparison of different protocols.

| protocol | $m_u$ | $q_k$ | $b_k$ | $\eta_1(\%)$ | $\eta_2(\%)$ |
|---|---|---|---|---|---|
| Dong et al. | 1 | 4 | 4 | 12.5 | 25 |
| Kao et al. | 1 | 4 | 4 | 12.5 | 25 |
| Proposed protocol | 2 | 6 | 3 | 22.22 | 33.33 |

## 6. Conclusions

In the present scheme, three-party quantum secure direct communication protocol based on three-particle GHZ-like state is proposed. In order to transmit two-bit secret message, entanglement swapping is utilized. Therefore, no qubits carrying the secret messages are transmitted. Also, this protocol is unconditionally secure, if the perfect quantum channel is used.

In this scheme, after sharing three-particle GHZ-like state between users, Alice encodes her two-bit secret messages by applying defined unitary operations on her particles. Also, the controller applies Hadamard operation on his particle. After that, Charlie and Alice use Bell measurement on their particles and announce the results outcomes to the receiver. Then, the receiver performs unitary operation on his particle according to the classical bits that he received. Then, he can find out the secret messages by measuring his particle. In addition, if the controller does not cooperate with the others communicators, they cannot have a successful communication. Maybe, the receiver cannot distinguish the correct information. By this protocol, the efficiency is improved compared with previous works.

In the future, we want to extend this protocol to a controlled bidirectional quantum secure direct communication. In other word, the two legitimate users can exchange their secret messages under the permission of the controller, simultaneously.